%% file: main.tex
\documentclass[fleqn,10pt]{wlscirep}
\usepackage{array}
\usepackage{graphicx}
\usepackage{colortbl}
\usepackage[utf8]{inputenc}
\usepackage[T1]{fontenc}
\usepackage{lineno}
\usepackage{physics}
\usepackage{amsmath}
\usepackage{todonotes}
\usepackage{longtable}
\usepackage{float}

\title{A Quantum-Chemical Bonding Database for Solid-State Materials}

\author[1,3]{Aakash Ashok Naik}
\author[1]{Christina Ertural}
\author[1]{Nidal Dhamrait}
\author[2]{Philipp Benner}
\author[1,3]{Janine George}
\affil[1]{Federal Institute for Materials Research and Testing, Department Materials Chemistry, Berlin, 12205, Germany}
\affil[2]{Federal Institute for Materials Research and Testing, eScience Group, Berlin, 12205, Germany}
\affil[3]{Friedrich Schiller University Jena, Institute of Condensed Matter Theory and Solid-State Optics, Jena, 07743, Germany}

\affil[*]{corresponding author(s): Janine George (janine.george@bam.de)}

\begin{abstract}
 An in-depth insight into the chemistry and nature of the individual chemical bonds is essential for understanding materials. Bonding analysis is thus expected to provide important features for large-scale data analysis and machine learning of material properties. Such chemical bonding information can be computed using the LOBSTER software package, which post-processes modern density functional theory data by projecting the plane wave-based wave functions onto a local, atomic orbital basis. With the help of a fully automatic workflow, the VASP and LOBSTER software packages are used to generate the data. We then perform bonding analyses on 1520 compounds (insulators and semiconductors) and provide the results as a database. The bonding analysis database structure is also explained, which allows easy data retrieval. The projected densities of states and bonding indicators are benchmarked on standard density-functional theory computations and available heuristics, respectively. Lastly, we illustrate the predictive power of bonding descriptors by constructing a machine-learning model for phononic properties, which shows an increase in prediction accuracies by ~27 \% (mean absolute errors) compared to a benchmark model differing only by not relying on any quantum-chemical bonding features. 
\end{abstract}
\begin{document}

\flushbottom
\maketitle

\thispagestyle{empty}

\section*{Background \& Summary}

Understanding the interactions between constituent atoms in crystalline materials paves the way for developing and tailoring novel solid-state materials with desired application-specific properties.\cite{hoffmann1987chemistry,albright2013orbital,burdett1995chemical,nelson2022chemical} For instance, the ultra-low lattice thermal conductivity in thermoelectric materials is connected to strong antibonding interactions.\cite{das2023strong,he2022accelerated} Bonding analysis aids in quantifying such interatomic interactions, and several theoretical frameworks exist. Popular and well-known approaches are the Atoms In Molecules (AIM) approach to derive electron density-based Bader charges,\cite{bader1981quantum} or wave function-based concepts like the Mulliken population analysis,\cite{mulliken1955electronic} from which Crystal Orbital Overlap Populations (COOP),\cite{hughbanks1983chains} Crystal Orbital Hamiltonian Populations (COHP)\cite{dronskowski1993crystal}, and the Crystal Orbital Bond Index (COBI)\cite{2021crystal} are derived.

Nowadays, many robust automation frameworks for simulation have become available.\cite{curtarolo2012aflow, pizzi2016aiida, mathew2017atomate, gjerding2021atomic, george_automation_2021}  These automation tools allow for high-throughput calculations on thousands of materials.\cite{toher_high-throughput_2014, de_jong_charting_2015,  petretto2018high} Reusing such large amounts of data as inputs for machine learning algorithms has enabled data-driven material science research for accelerated discovery of novel materials and gaining a better understanding between materials structure and properties.\cite{hautier_finding_2019, he2022accelerated}

For solid-state materials, plane wave-based basis sets provide easy means to exploit periodicity and gain computational efficiency due to their delocalized nature when performing atomistic simulations via density functional theory (DFT). This computational efficiency comes at the cost of losing crucial atom-specific chemical bonding information. The Local Orbital Basis Suite Toward Electronic-Structure Reconstruction (LOBSTER)\cite{deringer2011crystal,maintz2013analytic,maintz2016lobster,nelson2020lobster} software package can recover such bonding information by projecting plane-wave-based wave functions onto atomic orbitals. Since its first release, this program has been used extensively to study  different materials classes (e.g, phase-change materials,\cite{konze2019exploring,hempelmann2022orbital}, Li/Na ion battery,\cite{HuangLiNanoporous,ertural2022first} low thermal conductivity materials\cite{hu2023mechanism,das2023strong,dutta2019bonding,sun2019achieving}) and to uncover the diverse underlying atomistic phenomena in the respective bonding mechanisms.\cite{sun2019achieving,hempelmann2022orbital,ertural2022first} Although high-throughput materials design and research studies with LOBSTER data have been conducted in a few cases,\cite{xi2018discovery,ohmer2019high,chanussot2021open,chanussot2021correction} no dedicated database exists to retrieve and reuse such data. Previous studies have clearly shown that bonding data computed with LOBSTER is of high value  for the materials informatics community, and we provide an open-access database of bonding information here for the first time. 

In this work, we perform bonding analysis for 1520 compounds using an automated workflow\cite{george2022automated} recently developed by some of us that combine Vienna Ab initio Simulation Package (VASP)\cite{kresse1996efficient,kresse1996efficiency,kresse1993ab} DFT computations with LOBSTER calculations using Python tools like \emph{pymatgen},\cite{ong2013python} \emph{atomate},\cite{mathew2017atomate} and \emph{FireWorks}.\cite{jain2015fireworks} To generate summarized bonding information ready to be used for machine learning studies, we used the \emph{LobsterPy}\cite{george2022automated,lobsterpy_zenodo} package that automatically analyzes LOBSTER COHP output files. We provide this summarized bonding information data as (lightweight) JSON files. We also distribute all relevant LOBSTER computation data validated and formatted using a \emph{Pydantic} schema, including all the settings and relevant output files. 

In the following sections, we begin by briefly summarizing the computational details of the workflow employed to perform the computations. We then describe the method used to generate entries in the database and provide an overview of the structure of the database. Finally, we benchmark the quality of our results by comparing them with projected densities of states from a widely-used density-functional theory code and available heuristics for bond valences and coordination environments. Lastly, we demonstrate the influence of including quantum-chemical bonding data in a machine-learned model for predicting phononic properties.

\section*{Methods}
\textbf{Structures.} We included a total of 1520 crystalline materials in this work. The Materials Project (MP) database\cite{jain2013commentary} is used to retrieve all the structures. These materials belong to a previously published dataset of harmonic phonon properties including band structures and densities of states.\cite{petretto2018high} We selected this database as it consists only of semiconductors and insulators. For these materials, it is easier to choose a local basis set for the LOBSTER projection as they have clearly distinguished valence and conducting states separated by a band gap. We chose a minimal basis consisting only of occupied valence orbitals in the atomic ground state of each atom (as used in the projector-augmented wave method).

\hfill\break
\noindent\textbf{Bonding indicators definitions.} LOBSTER first projects the projector-augmented wave (PAW) wavefunctions obtained from DFT computations onto a local orbital basis to quantify the interatomic interactions. Combining the coefficients of linear combinations of atomic orbitals (LCAO) generated from this projection with overlap, Hamiltonian, and density matrices, quantum-chemical bonding characteristics in materials are estimated. Here, we summarize the key quantities computed by LOBSTER, and the notations used follow the same convention as in Ref.~\citenum{2021crystal}:\cite{nelson2020lobster,2021crystal}

\begin{gather}
\operatorname{pCOOP}_{\mu \nu}(E) = S_{\mu \nu} \sum_{j, k} w_k \operatorname{Re}\left(c_{\mu, j k}^* c_{\nu, j k}\right) \cdot \delta\left(\varepsilon_j(\boldsymbol{k})-E\right) \\
\operatorname{pCOHP}_{\mu \nu}(E) = H_{\mu \nu} \sum_{j, k} w_k \operatorname{Re}\left(c_{\mu, j k}^* c_{\nu, j k}\right) \cdot \delta\left(\varepsilon_j(\boldsymbol{k})-E\right) \\
\operatorname{COBI}_{\mu \nu}(E) = P_{\mu \nu} \sum_{j, k} w_k \operatorname{Re}\left(c_{\mu, j k}^* c_{\nu, j k}\right) \cdot \delta\left(\varepsilon_j(\boldsymbol{k})-E\right)
\end{gather}
The overlap, Hamiltonian and density matrix between orbitals $\Phi_{\mu}$ and $\Phi_{\nu}$ are represented by $S_{\mu \nu}$, $H_{\mu \nu}$ and $P_{\mu \nu}$ respectively. $w_k$ is the $k$-point weight, and $c_{\mu, j k;\nu, j k}$ are the coefficients of LCAOs. $\operatorname{Re}$ indicates the real part of the complex value. $\varepsilon_j(k)$ and $E$ represent the energy eigenvalue of band $j$ at $k$ within the Brillouin zone and the general energy, respectively. The energy-integrated values (up to the Fermi level) of these quantities, namely ICOOP, ICOHP, and ICOBI, can be interpreted as the number of electrons in the bond, a measure of bond covalency (corresponding to bond strength), and bond order, respectively.

LOBSTER also provides Mulliken and L\"owdin atomic charges from the orbital-derived atomic gross populations (GP).\cite{ertural2019development} The Madelung energy is derived using Mulliken or L\"owdin atomic charges as input. Madelung energies represent the electrostatic part of the lattice energy and can be related to the stability of ionic crystal structures. For details about the mathematical formulation related to Madelung energies, Mulliken, and  L\"owdin atomic charges in LOBSTER, we refer the readers to Ref.~\citenum{2021crystal} and the literature referenced therein.

\begin{figure}[h]
\centering
\includegraphics[width=.90\linewidth]{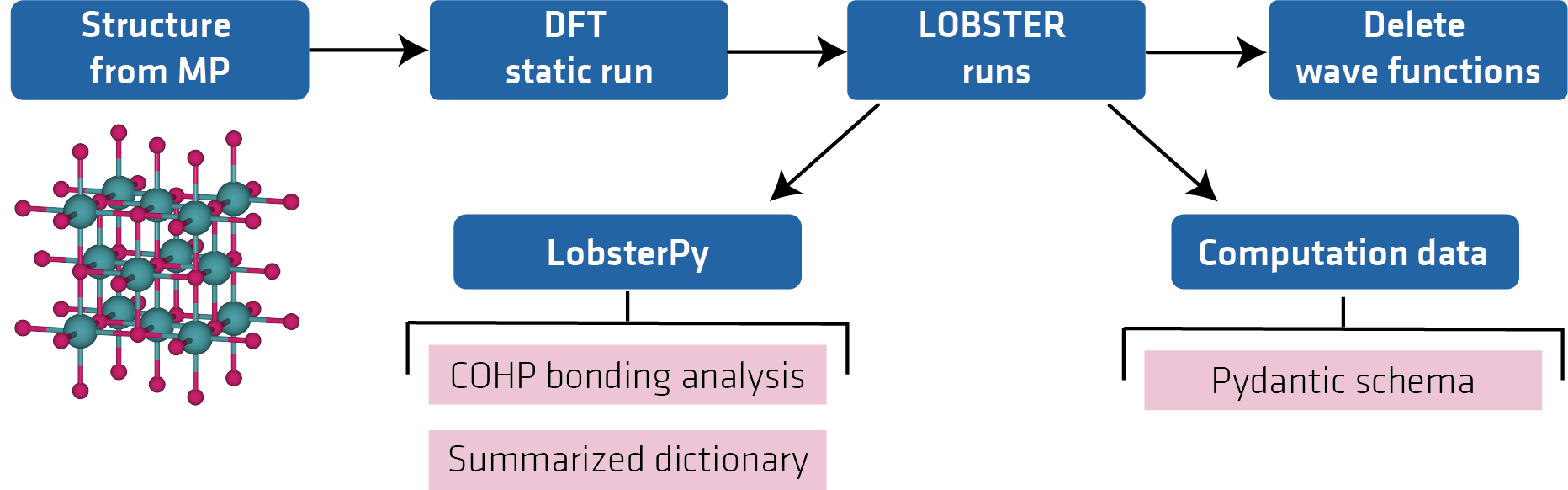}
\caption{Workflow schematic for computations and data record generation.}
\label{fig:Fig1}
\end{figure}

\hfill\break
\noindent\textbf{Workflow and computational parameters.} To create the database, we used an automatic bonding analysis workflow\cite{george2022automated} developed recently by some of us. To start this workflow, one must provide the crystal structure as input. Based on the input structure, it performs the bonding analysis with the LOBSTER\cite{nelson2020lobster} program by adding all necessary computational steps to the pipeline. To summarize, these steps involve (a) writing VASP input files with an appropriate number of bands (NBANDS) for a static DFT run, (b) a static DFT run, (c) writing input files for LOBSTER runs with all available atomic orbital basis functions for the projection of the wave function, (d) LOBSTER runs, (e) deleting (disk-space consuming) wave function (WAVECAR) files. Fig.~\ref{fig:Fig1} shows the schematic sequence of our workflow. 

Within this workflow, the DFT computations were performed using the generalized gradient approximation (GGA) functional as parameterized by Perdew, Burke, and Ernzerhof (PBE)\cite{perdew1992atoms,perdew1996generalized} within the PAW framework.\cite{kresse1999ultrasoft, blochl1994projector}
We employ a grid density of 6000 $k$-points per reciprocal atom and set NEDOS (number of energy points on which the density of states is evaluated) to 10000 points. The electronic structure's convergence criterion is set to $10^{-6}$ eV, and the plane-wave energy cutoff is set to the standard value of 520 eV, as implemented in the original workflow. The Brillouin zone is integrated using the tetrahedron method with Blöchl correction\cite{blochl1994improved} (i.e., ISMEAR=-5). All computations were performed including spin polarization. For COHP computations using LOBSTER, we use the entire energy range of VASP static runs, and COHP steps are set equal to the NEDOS (i.e., 10000 steps) set for VASP static run. We increased the number of points for the DOS computation to be able to benchmark the LOBSTER projected DOS with the help of the VASP projected DOS. As both LOBSTER and VASP DOS were computed in the same workflow, the VASP DOS was also computed without symmetry (ISYM=0), which is now also the recommended setting for VASP projected DOS for the VASP version that we used.\cite{waybackvasp}  With this high number of points in the DOS and COHP computations, the bonding and anti-bonding percentage values from our automatic analysis of output files additionally also pose a very good estimate of bonding and anti-bonding contribution in bonds as we rely on a numerical integration in LobsterPy. The code for starting the workflows is also provided for reproducibility.

\hfill\break
\textbf{Generating data records.} We provide data records in two forms. The smaller data record consists of summarized bonding information that is very lightweight and can be quickly assessed in seconds to retrieve and examine relevant bonds. The other, larger data record consists of all the LOBSTER computational data. 

To generate the smaller data records including summarized bonding information (LOBSTER lightweight data), we used the CondensedBondingAnalysis schema implemented as part of the \emph{atomate2~}\cite{waybackatomate2}LOBSTER workflow. This schema automatically analyzes the LOBSTER output files in the ``cation-anion'' and ``all'' bond modes using the \emph{LobsterPy}\cite{george2022automated,lobsterpy_zenodo} package. In cases without ions in the structure, only data from the analysis of all bonds are available.  When the "cation-anion" mode is used, the automatic analysis detects cations and anions based on the Mulliken charges, and only "cation-anion" bonds are included in the analysis. Then, the strongest cation-anion bond is determined based on the Integrated Crystal Orbital Hamiltonian Populations (ICOHPs). To determine coordination environments and to perform automatic plots, only bonds with a strength of at least 10\% of the strongest bond are considered. If the "all" mode is used, the other bonds are also included in the analysis. In addition, the schema identifies the strongest bonds and corresponding bond lengths based on ICOHP, ICOOP, and ICOBI data for the relevant bond pairs as per \emph{LobsterPy} bonding analysis. Additionally, we include Madelung energies and atomic charges based on Mulliken and L{\"o}wdin population analysis methods. A larger data record (Computational data) with all the important LOBSTER computation data is generated using the LobsterTaskDocument, which is a \emph{pydantic} schema again implemented as part of the \emph{atomate2} LOBSTER workflow.  This schema uses LOBSTER parsers implemented in the \emph{pymatgen} package to read the LOBSTER files and store the information necessary to recreate the Python objects in the form of a Python dictionary. It also includes all the data from smaller summarized bonding information data records. A code to generate and read these JSON files is also provided in the code repository for this publication. This allows easy means to reuse or access the data.

\vspace{-.35cm}
\section*{Data Records}
\textbf{LOBSTER lightweight data file format:}
\begin{table}[h]
\centering
\caption{Top level keys of the LOBSTER lightweight JSON files}
\label{tab:lightweightjsonrootkeys}
\resizebox{.97\linewidth}{!}{%
\begin{tabular}{|>{\hspace{0pt}}m{0.18\linewidth}|>{\hspace{0pt}}m{0.09\linewidth}|>{\hspace{0pt}}m{0.76\linewidth}|} 
\hline
\rowcolor[rgb]{0.161,0.522,0.941} Root keys & \multicolumn{1}{>{\centering\hspace{0pt}}m{0.09\linewidth}|}{Datatype} & Description \\ 
\hline
all\_bonds & dict & Summarized relevant bonds data (See table 2 for details)\\ 
\hline
cation\_anion\_bonds & dict & Summarized relevant cation-anion bonds data (See table 2 for details) \\ 
\hline
madelung\_energies & dict  & Total electrostatic energy for the structure as calculated from the Mulliken and L{\"o}wdin charges \\ 
\hline
charges & dict & Atomic charges with Mulliken and L{\"o}wdin population analysis methods as keys. Each key's corresponding list follows the order of sites in the crystal structure.\\ 
\hline
\end{tabular}
}
\end{table}
The data is stored in JSON format (See Data Citation\cite{zenodo_part1}). The files are named with the the Materials Project ID of the compound. Each JSON file includes summarized bonding information. Table.~\ref{tab:lightweightjsonrootkeys} summarizes the root keys to access data from the JSON file.Table~\ref{tab:innerkeys}, explains the data inside the ``all\_bonds'' and ``cation\_anion\_bonds'' keys.
\begin{table}[h]
\centering
\caption{Keys corresponding to ``all\_bonds'' and ``cation\_anion\_bonds'' in LOBSTER lightweight data JSON file}
\label{tab:innerkeys}
\resizebox{.97\linewidth}{!}{%
\begin{tabular}{|>{\hspace{0pt}}m{0.25\linewidth}|>{\hspace{0pt}}m{0.09\linewidth}|>{\hspace{0pt}}m{0.69\linewidth}|} 
\hline
\rowcolor[rgb]{0.161,0.522,0.941} Root keys & \multicolumn{1}{>{\centering\hspace{0pt}}m{0.09\linewidth}|}{Datatype} & Description \\ 
\hline
lobsterpy\_data & dict & Condensed bonding analysis data from LobsterPy (See table 3 for details) \\ 
\hline
lobsterpy\_text & string & Contains LobsterPy automatic analysis summary text \\ 
\hline
sb\_icobi & dict  & Dict with the strongest ICOBI bonds \\ 
\hline
sb\_icohp & dict & Dict with the strongest ICOHP bonds \\ 
\hline
sb\_icoop & dict & Dict with the strongest ICOOP bonds\\ 
\hline
\end{tabular}
}
\end{table}

\begin{table}[H]
\centering
\caption{Keys corresponding to ``lobsterpy\_data'' in LOBSTER lightweight data JSON file}
\label{tab:lobsterpykeys}
\resizebox{.97\linewidth}{!}{%
\begin{tabular}{|>{\hspace{0pt}}m{0.25\linewidth}|>{\hspace{0pt}}m{0.09\linewidth}|>{\hspace{0pt}}m{0.69\linewidth}|} 
\hline
\rowcolor[rgb]{0.161,0.522,0.941} Root keys & \multicolumn{1}{>{\centering\hspace{0pt}}m{0.09\linewidth}|}{Datatype} & Description \\ 
\hline
formula & string & Chemical formula of the compound \\ 
\hline
max\_considered\_bond\_length & float & Maximum bond length that has been considered in the analysis~ \\ 
\hline
limit\_icohp & float array  & Minimum and maximum ICOHP that has been considered in the analysis \\ 
\hline
number\_of\_considered\_ions & int & Number of ions that has been detected \\ 
\hline
sites & string & Site index of the sites in the crystal structure for which bonds have been detected (nested dict that describes the bond and its co-ordination environment as determined based on the ICOHP values. ) \\ 
\hline
type\_charges & string & Whether the Mulliken or the Löwdin charges have been used for the bonding analysis. \\ 
\hline
cuttoff\_icohp & float & ICOHP cutoff value set for bonding analysis\\
\hline
summed\_spins & bool & Indicates if spins are summed\\
\hline
start & int & Sets the energy for evaluation of bonding and anti-bonding percentages based on COHP \\
\hline
cohp\_plot\_data & dict & Relevant bond labels as keys and corresponding cohp objects to plot COHP curves from automatic analysis\\
\hline
which\_bonds & string & Indicates the mode of automatic bonding analysis run. (``cation\_anion'' or ``all'')\\
\hline
final\_dict\_bonds & dict & Includes relevant bond label, ICOHP mean value and indicates if anti-bonding states below the Fermi level exists\\
\hline
final\_dict\_ions & dict & Includes all different coordination environments and counts them\\
\hline
run\_time & float & Time needed in secs to run the automatic bonding analysis.  \\
\hline
\end{tabular}
}
\end{table}
\hfill\break
\textbf{Computational data file format:} The data is stored in JSON format (See Data Citations\cite{zenodo_part1,zenodo_part2}). The files are named as per the Materials Project ID of the compound. Each JSON file includes all the LOBSTER output files parsed and stored in the form of a Python dictionary. It also includes the summarized bonding analysis based on ICOHP values and contains the same information as explained in Table~\ref{tab:innerkeys}. Table.~\ref{tab:computationaldata} summarizes root keys to access data from the JSON file.
\begin{longtable}[c]{|l|l|l|}
\hline
\rowcolor[HTML]{2985F0} 
Root Keys & Data type & Description \\ \hline
\endfirsthead
\multicolumn{3}{c}%
{{\bfseries Table \thetable\ continued from previous page}} \\
\hline
\rowcolor[HTML]{2985F0} 
Root Keys & Data type & Description \\ \hline
\endhead
structure & dict & \begin{tabular}[c]{@{}l@{}}Dict representation of the pymatgen Structure object\\ used for the LOBSTER calculation\end{tabular} \\ \hline
charges & dict & \begin{tabular}[c]{@{}l@{}}Atomic charges dict from LOBSTER based on \\ the Mulliken and L{\"o}wdin charge analysis\end{tabular} \\ \hline
lobsterin & dict & LOBSTER calculation inputs \\ \hline
lobsterout & dict & Information on LOBSTER calculation output \\ \hline
lobsterpy\_data & dict & \begin{tabular}[c]{@{}l@{}}Summarized bonding analysis data from LobsterPy \\ (all bonds mode). It also includes Cohp objects \\ to plot the COHP curves from the automatic analysis\end{tabular} \\ \hline
lobsterpy\_text & dict & LobsterPy automatic analysis summary text (all bonds mode) \\ \hline
strongest\_bonds\_icohp & dict & Describes the strongest ICOHP bonds \\ \hline
strongest\_bonds\_icoop & dict & Describes the strongest ICOOP bonds \\ \hline
strongest\_bonds\_icobi & dict & Describes the strongest ICOBI bonds \\ \hline
lobsterpy\_data\_cation\_anion & dict & \begin{tabular}[c]{@{}l@{}}Summarized bonding analysis data from LobsterPy \\ (cation-anion bonds mode). It also includes Cohp objects \\ to plot the COHP curves from the automatic analysis\end{tabular} \\ \hline
lobsterpy\_text\_cation\_anion & dict & \begin{tabular}[c]{@{}l@{}}LobsterPy automatic analysis summary text \\ (cation-anion bonds mode)\end{tabular} \\ \hline
strongest\_bonds\_icohp\_cation\_anion & dict & Describes the strongest cation-anion ICOHP bonds \\ \hline
strongest\_bonds\_icoop\_cation\_anion & dict & Describes the strongest cation-anion ICOOP bonds \\ \hline
strongest\_bonds\_icobi\_cation\_anion & dict & Describes the strongest cation-anion ICOBI bonds \\ \hline
cohp\_data & dict & \begin{tabular}[c]{@{}l@{}}Dict representation of pymatgen CompleteCohp object \\ including data to plot COHP curves\end{tabular} \\ \hline
coop\_data & dict & \begin{tabular}[c]{@{}l@{}}Dict representation of pymatgen CompleteCohp object \\ including data to plot COOP curves\end{tabular} \\ \hline
cobi\_data & dict & \begin{tabular}[c]{@{}l@{}}Dict representation of pymatgen CompleteCohp object \\ including data to plot COBI curves\end{tabular} \\ \hline
dos & dict & \begin{tabular}[c]{@{}l@{}}Dict representation of pymatgen LobsterCompleteDos object \\ including the DOSCAR.lobster data\end{tabular} \\ \hline
lso\_dos & dict & \begin{tabular}[c]{@{}l@{}}Dict representation of pymatgen LobsterCompleteDos object \\ including the DOSCAR.LSO.lobster data\end{tabular} \\ \hline
madelung\_energies & dict & \begin{tabular}[c]{@{}l@{}}Consists of the Madelung energies of the structure derived from \\ the Mulliken and L{\"o}wdin charges\end{tabular} \\ \hline
\caption{Top level keys of computational data JSON files}
\label{tab:computationaldata}\\
\end{longtable}

\section*{Technical Validation}

\textbf{Projection quality}\hfill\break The absolute charge spilling reported at the end of the LOBSTER calculations indicates the quality of the projection corresponding to the loss of charge density that occurs when projecting the original PAW functions onto the local basis. Ideally, when the provided local basis set is complete (i.e., properly reproducing the PAW-based Hilbert space and representing the chemistry of the compound in question), the charge spilling value approaches zero, indicating the reliability of the results. Fig.~\ref{fig:Fig2} below shows the distribution of the charge spilling for our data set. Approximately 99\,\% of compounds have charge spilling of <~5\,\% .
\begin{figure}[H]
\centering
\includegraphics[width=.75\linewidth]{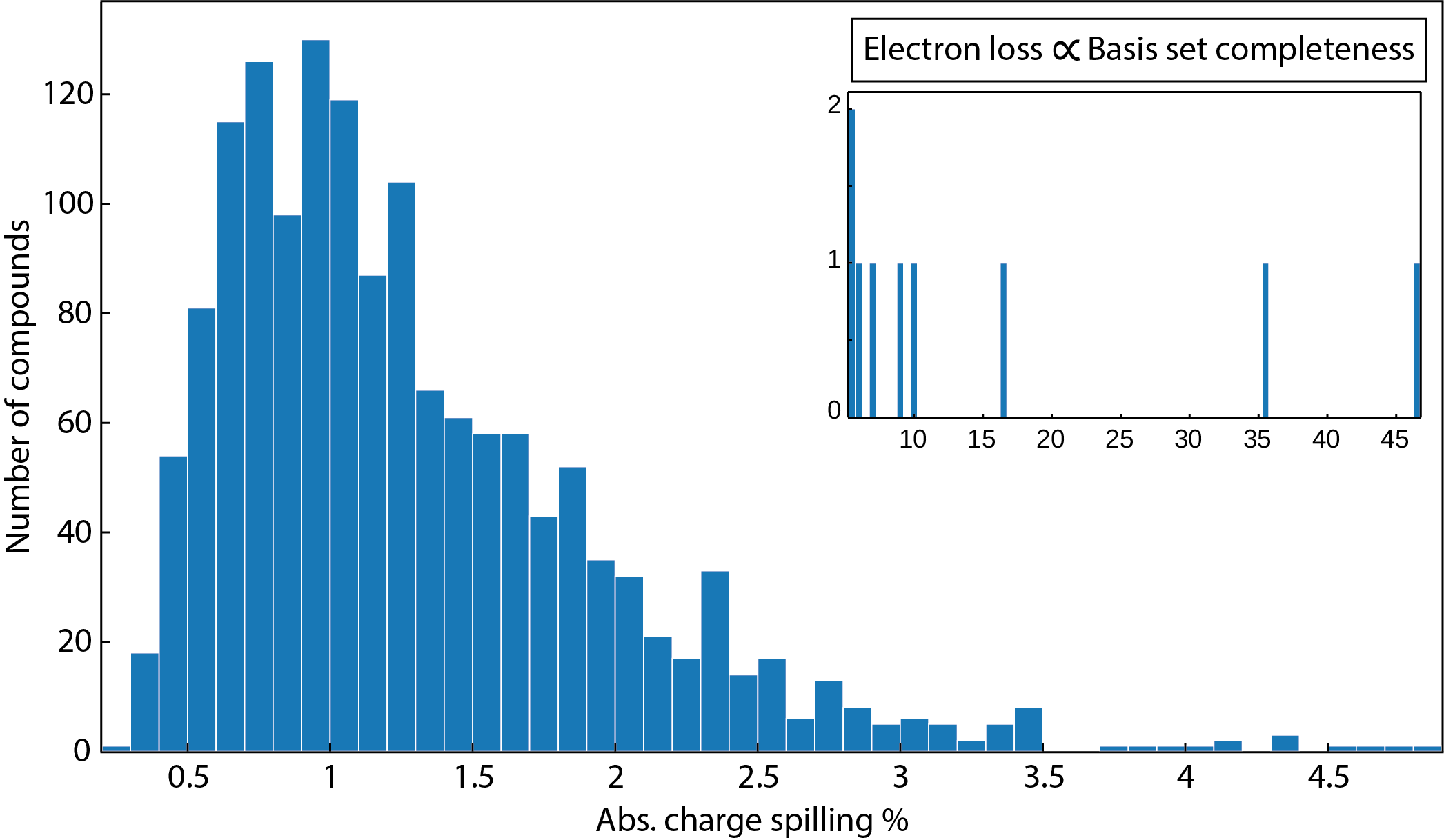}
\caption{Distribution of absolute charge spilling from LOBSTER computations for the entire data set (spilling > 5\,\% shown in the inset). Possible reasons for the nine outliers are discussed in the text.}
\label{fig:Fig2}
\end{figure} 
\noindent Only a very few compounds show a charge spilling of >~5\,\%, possibly due to the limited basis function availability in LOBSTER. 
The nine compounds showing an absolute charge spilling >~5\,\% are BaO$_2$ (mp-1105), SiC (mp-11713), Be$_2$C (mp-1569), Li$_4$NCl (mp-29149), CsBiO$_2$ (mp-29506), Cs$_2$O (mp-7988), KYO$_2$ (mp-8409), Rb$_2$PtSe$_2$ (mp-8622) and SrHfN$_2$ (mp-9383), with spillings ranging between 5.5 and almost 50\,\% (see inset in Fig.~\ref{fig:Fig2}). The most extreme case is BaO$_2$ with an absolute charge of 46.7\,\%. Two possible reasons for this outlier are coming into consideration: either the structure from the Materials Project database is not optimally relaxed, or the provided basis functions are not sufficient for a proper projection. In the first case, an additional optimization of the MP structure leads to an absolute charge spilling of 3.91\,\%. In the second case, with an experimental version of LOBSTER,\cite{ErturalDiss} that allows to include arbitrary orbitals into the projection, adding the La 5d orbital to Ba, as the VASP POTCAR suggests a 5d occupation of 0.010, the absolute charge spilling drops to 1.40\,\% without further structural optimization. We have included this compound in the rest of the analysis and in the database, as the other benchmarked results still show sufficient agreement. However, the bonding information from this compound should be used with caution.

Overall, these results demonstrate that the local basis used for our computations correctly represents the material's chemistry for the majority of compounds. The LOBSTER projection mismatch (abs. charge spilling > 5\,\%) also helps to point out possible inconsistencies in the Materials Project database, such as not fully optimized structures, or to suggest further improvements for VASP pseudopotentials and LOBSTER basis functions as these may be an error source, as discussed in the case of BaO$_2$.

\hfill\break
\noindent\textbf{Projected density of states (PDOS) benchmarking}\hfill\break
\noindent As LOBSTER quantifies the inter-atomic interactions by projecting the PAW wavefunctions from DFT computations (in our case: VASP) onto a provided local orbital basis, it also generates PDOS that is independent of PDOS generated by VASP. But unlike the LOBSTER projection, the VASP projection typically loses more electron density when using standard Wigner-Seitz radii. Nevertheless, we will use the VASP projection data for benchmarking as this data is commonly used in the field, and automation are available. We will, however, not compare the absolute projected density of state values for this reason. A common way to compare the density of states relies on visual inspection of relevant features. However, with thousands of PDOS plots, performing a visual inspection is not feasible. To numerically compare the PDOS from VASP and LOBSTER, we have chosen several methods that do not rely on the absolute values but instead on features of the PDOS that are relevant for understanding the electronic structure of a material. First, we compute moments of the PDOS from VASP and LOBSTER. These moments, in principle, provide an estimate of the shape of the PDOS in the selected energy range. Namely, we compare here the band center ($1^{st}$ moment),\cite{hammer1995electronic} bandwidth (the $\sqrt{2^{nd}}$  moment), band skewness (the $3^{rd}$ standardized moment), and kurtosis (the $4^{th}$ standardized moment) of the band directly below the Fermi level ($E_F$). These features provide an overview of the numerical similarity of DOS and are easy to evaluate using existing methods implemented in electronic\_structure.dos module in \emph{pymatgen}.\cite{ong2013python,rosen2023free} It must be noted that we compare the  L{\"o}wdin symmetric orthonormalized (LSO) DOS obtained from LOBSTER, which recovers the entire Hilbert space and ensures that no electron density is lost, to the VASP projected DOS.
\begin{figure}[h]
  \begin{minipage}{.90\linewidth}
  \centering
  \includegraphics[width=\textwidth]{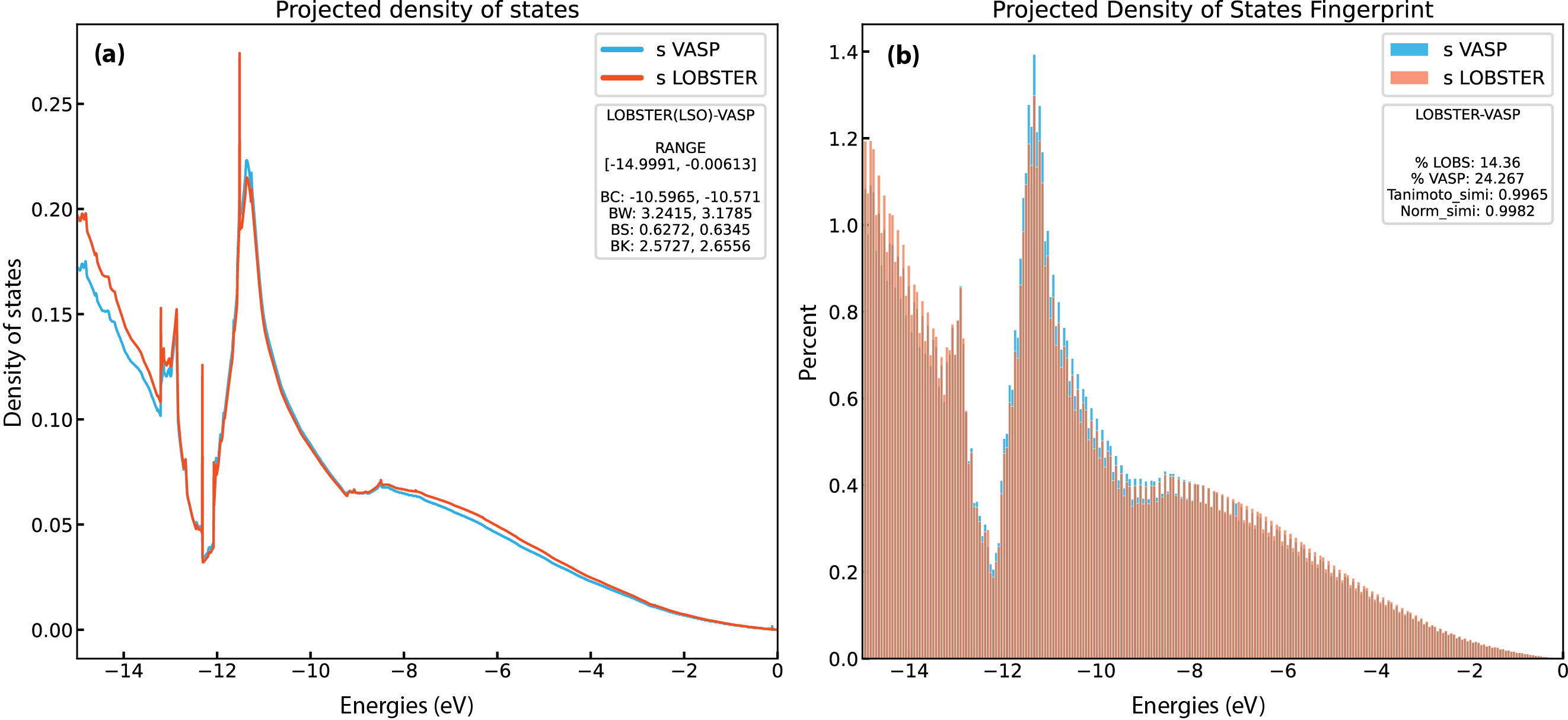}
  \end{minipage}
  \caption{(a) Band features and (b) Fingerprint exemplar plots for PDOS from LOBSTER and VASP runs for diamond (mp-66). In subfigure (a), BC, BW, BS, and BK denote band center, width, skewness, and kurtosis, respectively. The percentages of orbital contribution in the chosen energy range are shown in subfigure (b) as \% LOBS and \%VASP. The Tanimoto index and the normalized vector dot product, respectively, are denoted by the Tanimoto\_simi and Norm\_simi.}
 \label{fig:exemplar}
\end{figure}

To compute the PDOS features, we first extract all energy ranges below $E_F$ in which the PDOS is not equal or close to zero.\begin{figure}[H]
  \begin{minipage}{\linewidth}
  \centering
  \includegraphics[width=.90\textwidth]{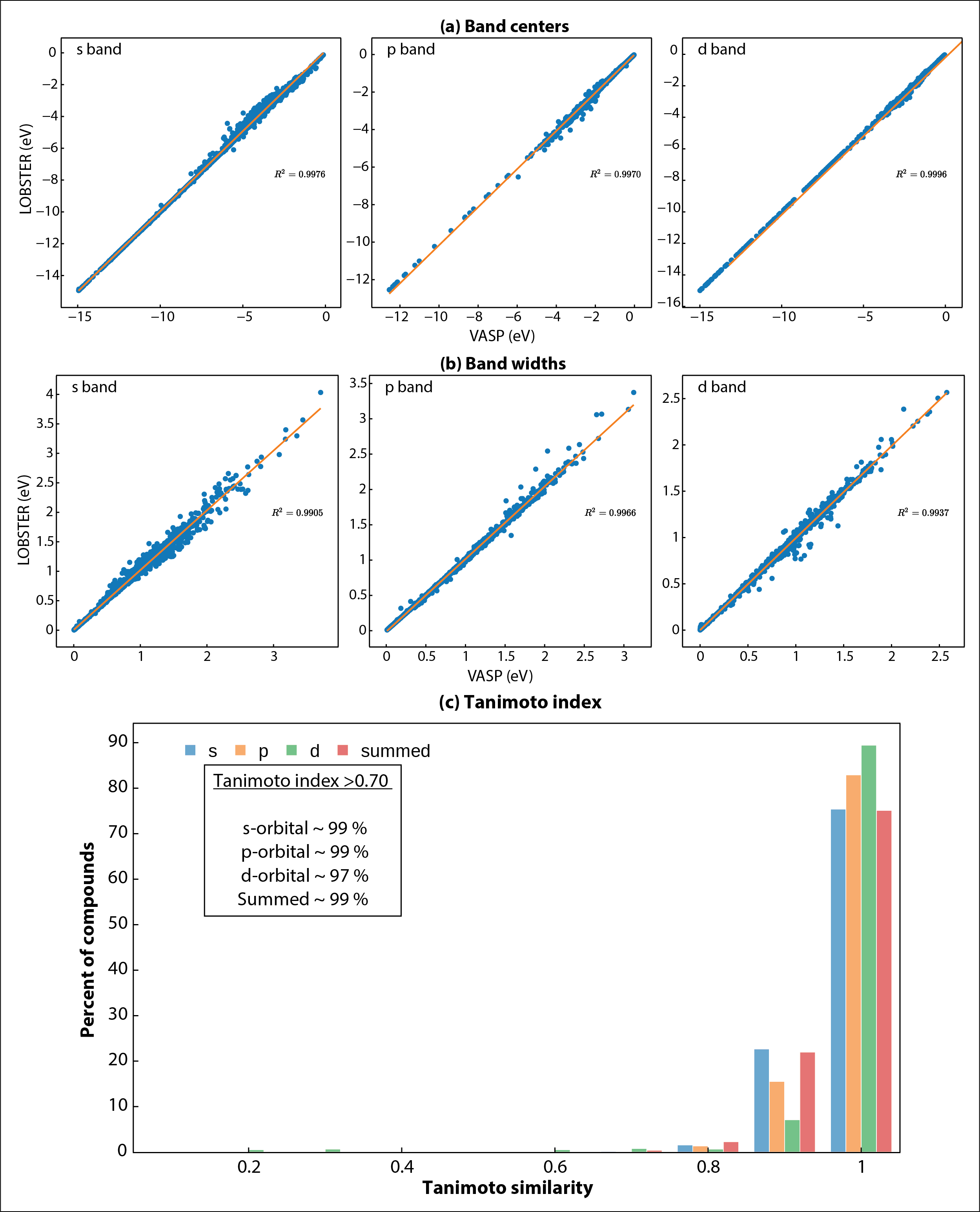}
  \end{minipage}
  \caption{(a) Band centers and (b) Band width comparison of projected DOS (s, p and d bands) for first energy range without PDOS values close or equal to zero below the Fermi level ($E_F$) obtained from LOBSTER and VASP runs. Both figures show that projected DOS from LOBSTER runs agree very well with our reference VASP data. (c) Histogram of Tanimoto index ($S_{A,B}$) computed between VASP and LOBSTER PDOS (Summed denotes the sum of all individual PDOS).}
 \label{fig:spdbandwidthcenter}
\end{figure}
\noindent  Next, we use the energy range just below $E_F$, where a non-zero PDOS is detected, to evaluate the PDOS moment features. To ensure that the obtained energy ranges significantly contribute to the overall band, we set a threshold of 0.5 electrons for the band feature comparisons. Fig.~\ref{fig:exemplar} (a) provides exemplar plots for comparing the PDOS. As evident from the band features, a sufficient agreement exists in this particular case (diamond, mp-66) between VASP and LOBSTER data. In Fig.~\ref{fig:spdbandwidthcenter} (a) and (b), we compare projected DOS for s, p, and d band centers and band widths obtained from our VASP and LOBSTER runs for the whole data set, respectively. A very good agreement is visible for most compounds. In Fig.~\ref{fig:spdbandskewnkurt}, we report comparisons of left-out PDOS features, namely band skewness and kurtosis. A comparison of the non-LSO DOS is also available in the in Fig.~\ref{fig:nonlsobf}.

Another way to assess the similarity between PDOS is to compute Tanimoto coefficients. Earlier studies have demonstrated that such a measure is not only suitable to compute the similarity between molecules\cite{bajusz2015tanimoto} but is also a reliable way to compare DOS of materials.\cite{kuban2022density} The formula to compute the Tanimoto coefficient is as follows:
\begin{equation}
S_{A,B}=\frac{A \cdot B}{||A||^2+||B||^2-A \cdot B}
\end{equation}
The Tanimoto coefficient ($S_{A,B}$) can be interpreted as the ratio of the dot product of the two vectors A and B to the sum of their magnitudes and the dissimilarity between them. We adapted the ``materials\_fp'' module of the FHI-vibes\cite{knoop2020fhi,knoop2020git} Python package to evaluate the similarity between the PDOS of the VASP and the LOBSTER program. The adapted code has been incorporated in the \emph{pymatgen} package and has been publicly available since v2023.1.9. Here, we first discretize PDOS from VASP and LOBSTER in 256 bins and normalize it before computing the $S_{A,B}$ for the energy range of $-15$ to $0$ eV (energies are shifted relative to the Fermi energy) for all the compounds. Again, for diamond (mp-66) in Fig.~\ref{fig:exemplar}, we show the binning of the PDOS and the corresponding Tanimoto similarity,  indicating very good agreement between VASP and LOBSTER data. Compounds, where the number of valence electrons obtained by integrating summed PDOS of VASP exceeded the actual valence electrons based on the POTCAR, are excluded from the analysis, as this indicates a poor projection. Again, we only compare PDOS if they significantly contribute to the density of states in the selected energy range. We have set this threshold to 5 \% of the sum of the projected DOS. Fig.~\ref{fig:spdbandwidthcenter} (c) shows the distribution of evaluated $S_{A,B}$ for the subset of our dataset. We can see that,  for most compounds, $S_{A,B}$ lies in the range of 0.75 to 1. Approximately 99\,\% of compounds have a similarity index of more than 0.70. Only a few cases exist where $S_{A,B}$ is less than 0.70, as shown in Fig.~\ref{fig:lowtanimoto}. Disagreements are observed in cases where unusual sharp peaks occur in the projection or some low-lying states are missing in VASP or LOBSTER projections. Overall our results demonstrate that the basic features of the PDOS from VASP and LOBSTER agree very well. Therefore, we can conclude that the LOBSTER projection was performed reliably and that we can compute bonding properties such as COHPs and COBIs of high quality based on this projection. We also provide an interactive dash app to explore these computed PDOS features visually for convenience (\href{https://doi.org/10.5281/zenodo.7795903}{10.5281/zenodo.7795903}).

\hfill\break
\noindent\textbf{Further quality markers: Atomic charges and coordination environments}\hfill\break While Mulliken and L\"owdin charges from LOBSTER are derived using the LCAO coefficients and arrive at non-integer values,\cite{ertural2019development} the bond valence analysis (BVA)\cite{o1991atom} derives classical integer oxidation states. To make these methods comparable, we chose to sample whether an atomic charge sign from the LOBSTER computations is positive or negative and compare it to the charge signs from the BVA method as implemented in \emph{pymatgen}. For the two approaches to agree, all constituent atoms in the crystal structure after one-to-one mapping must be classified the same way, i.e., as cations or anions. Here we see 96\%  agreement between LOBSTER's Mulliken charge analysis results and the BVA method. Deviations can be found in compounds having small electronegativity differences between the constituent atom pairs, i.e., for non-ionic compounds. Supplementary information Fig.~\ref{fig:endiff} shows the electronegativity difference between atom pairs for compounds where disagreement between BVA and Mulliken atom classification is observed. We highlight the elements where we encounter disagreement in red. A closer look at this figure reveals that a handful of intermetallic, M--H, M--P, and M--B interactions (involving semimetals) are mismatched. An overview of the involved elements is also given as a heatmap in Fig.~\ref{fig:heatmap}.

\emph{LobsterPy} can evaluate coordination environments directly based on the electronic structure by taking the ICOHP (a covalent bond strength measure) into account.\cite{george2022automated,waroquiers2020chemenv,pan2021benchmarking} The ICOHPs are used to determine the neighboring atoms. In this comparison, we only focus on bonds between cations and anions as determined by the Mulliken charges. Based on the shapes formed by the neighboring atoms, distances to ideal reference polyhedra are then used to determine the closest polyhedra.  To validate the coordination environments from \emph{LobsterPy}, we are benchmarking them with purely geometrically determined ones as determined by ChemEnv.\cite{waroquiers2020chemenv} In ChemEnv, multiple strategies are available to determine coordination environments. Here we use the SimplestChemEnvStrategy to determine the neighbors, which under the hood, uses a Voronoi partitioning scheme. We set the distance and solid-angle cutoffs to recommended values of 1.4 and 0.3, respectively. To only include cation-anion bonds, we again use the BVA method to determine the ideal oxidation states. Comparing the coordination environments detected for each site, we see an agreement for 79\,\% of the sites. Thus, the coordination environments from our database agree very well with those determined by commonly used geometric algorithms.
\hfill\break

\begin{figure}[h]
  \centering
  \includegraphics[width=\textwidth]{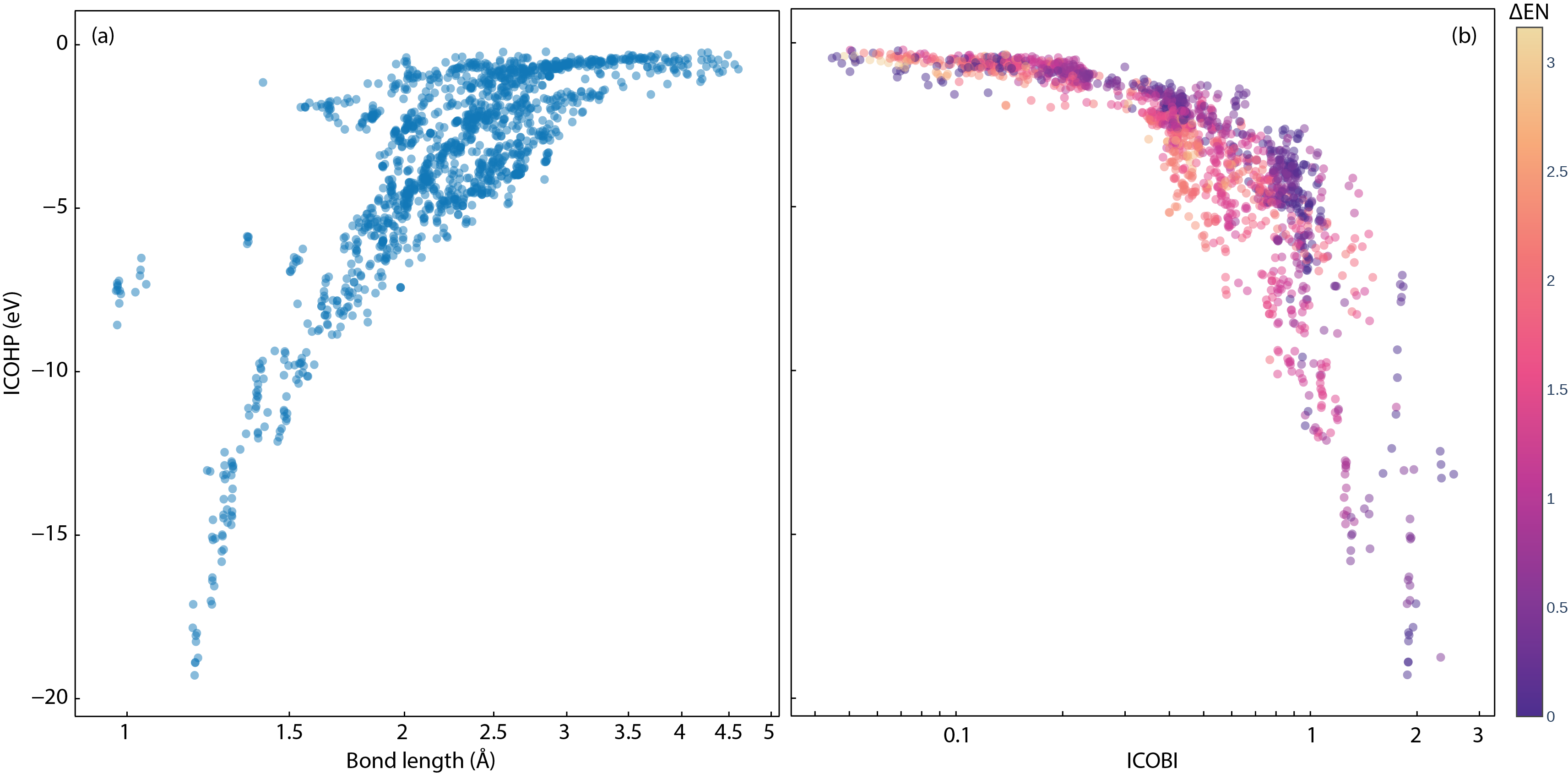}
  \caption{(a) The strongest ICOHP values for each compound and their respective bond lengths (b) Strongest ICOHP compared against two-center ICOBI interaction (logarithmic scale). Data points are colored according to the Pauling electronegativity difference between pairs of atoms.}
 \label{fig:lobsterplot}
\end{figure}

\noindent\textbf{Data exploration and utility}\hfill\break
First, we evaluate the bonding indicators in more detail. The most negative ICOHP value indicates the strongest covalent interaction per definition. Plotting the strongest ICOHP values (eV) found per compound and their corresponding bond lengths (\AA~) as shown in Fig.~\ref{fig:lobsterplot} (a), we see the expected decrease in covalent bond strengths with increasing bond lengths. In a bond range from about 1\,\AA~to 2\,\AA, a steep relation between ICOHP and bond distance can be observed, which eventually flattens for longer bond distances, indicating the short-ranged nature of covalency. The outliers around 1\,\AA~within the ICOHP energy range from $-$5 to $-$10\,eV are O--H and N--H bonds (cf. interactive plots: \href{https://doi.org/10.5281/zenodo.7802325}{10.5281/zenodo.7802325}). As covalent bonds between hydrogen and other nonmetal elements are known to be shorter and rather strong in nature,\cite{gordy1946relation,Benson1965iii,missong2015synthesis} this finding is no surprise. 

Fig.~\ref{fig:lobsterplot} (b) compares the strongest ICOHP and two-center ICOBI interactions for each compound from LOBSTER computations. Each data point is colored according to the Pauling electronegativity difference ($\Delta$EN) between the interacting atoms. More details can be found in the interactive plot (\href{https://doi.org/10.5281/zenodo.7802325}{10.5281/zenodo.7802325}). Up to a bond order (ICOBI) of 0.3 (weak bond range), the change of the ICOHP with growing ICOBI is smaller than after this value.  After that, the covalent bond strength increases rapidly with the bond order, demonstrating the different sensitivity of ICOHP and ICOBI with respect to changes in the chemical bonding environment.

Of course, the more ionic interactions (larger $\Delta$EN) can be found within the smaller ICOHP and ICOBI (absolute) values, as both descriptors indicate covalent interactions until eventually only interactions with small $\Delta$EN dominate for the interval ICOHP~<~$-7$\,eV and ICOBI~>~1. The interactions with very small (absolute) ICOHP and ICOBI values labeled as covalent according to $\Delta$EN are metal-metal (weak covalent) interactions like Rb--Rb or Rb--Cs contacts. 
Then there is a range of ICOHP (around $-$0.7 to $-$2.0\,eV) and ICOBI (around 0.25 to 0.45) values containing Zintl-like intermetallic phases like Na$_2$TlSb (mp-866132), RbAg$_3$Te$_2$ (mp-10481), KZnSb (mp-7438), KCuTe (mp-7436), Na$_2$AgSb (mp-7392), K$_2$AgSb (mp-7643), Na$_2$AgAs (mp-8411), K$_2$CuSb (mp-10381), K$_5$CuSb$_2$ (mp-27999), RbTeAu (mp-9008), K$_2$SbAu (mp-867335), KAuSe$_2$ (mp-29138) or Na$_2$AsAu (mp-7773) and more ($\Delta$EN for the respective bonds ranges between 0.1 and 0.5). This is particularly interesting since Zintl phases and related intermetallic compounds are of great interest for thermoelectric candidates\cite{B702266B,sun2019achieving} and, e.g., Na$_2$TlSb\cite{yue2023strong} and KCuTe\cite{C9RA07828B} show thermoelectric behavior. Phase-change and thermoelectric materials contain two-center interactions that tend to show smaller ICOHP and ICOBI values than expected from pure electronegativity differences as they are fragments of (hypervalent) multi-center bonds.\cite{2021crystal,nelson2022chemical, ErturalDiss, hempelmann2022orbital} In comparison to diamond (ICOHP = $-$9.6\,eV here and in ref.\,\citenum{nelson2022chemical}) and silver (ICOHP = $-$0.2\,eV from ref.\,\citenum{nelson2022chemical}) the two-center bond characteristic regarding the ICOHP lies between metallic and covalent bonding type (such as GeTe with ICOHP = $-$1.8\,eV in ref.\,\citenum{nelson2022chemical} and $\Delta$EN = 0.09) and is hence related to the meta-valent bonding mechanism.\cite{hempelmann2022orbital, lee2021multi,yu2020chalcogenide, pries2019phase,jones2022chemical} As we have only calculated semiconducting and insulating materials, a purely metallic bonding mechanism can be excluded. Chemically similar compounds in our data set with the classic relation between ICOHP and $\Delta$EN are, e.g., Rb$_3$BAs$_2$ (mp-9718, ICOHP(As--B) = $-$7.4\,eV, $\Delta$EN = 0.14), BSb (mp-997618, ICOHP = $-$5.0\,eV, $\Delta$EN = 0.01) and Ga$_2$Se$_3$ (mp-1340, ICOHP = $-$5.4\,eV, $\Delta$EN = 0.74). It needs to be proven if the relevant compounds from our data set exhibit multi-center ICOBI as well, as it would open up a way to use the ICOHP vs. ICOBI plot as a materials map\cite{nelson2022chemical, yu2020chalcogenide, pries2019phase,esser2017automated,schon2022classification} for thermoelectric (and phase-change) materials. In summary, we could demonstrate on a larger scale that ICOHP and ICOBI classify bonds according to covalency, and another indicator would be needed to further distinguish the weak covalent interactions as metallic, ionic, or (potential) multi-center interactions.
\begin{figure}[H]
  \centering
  \includegraphics[width=.95\textwidth]{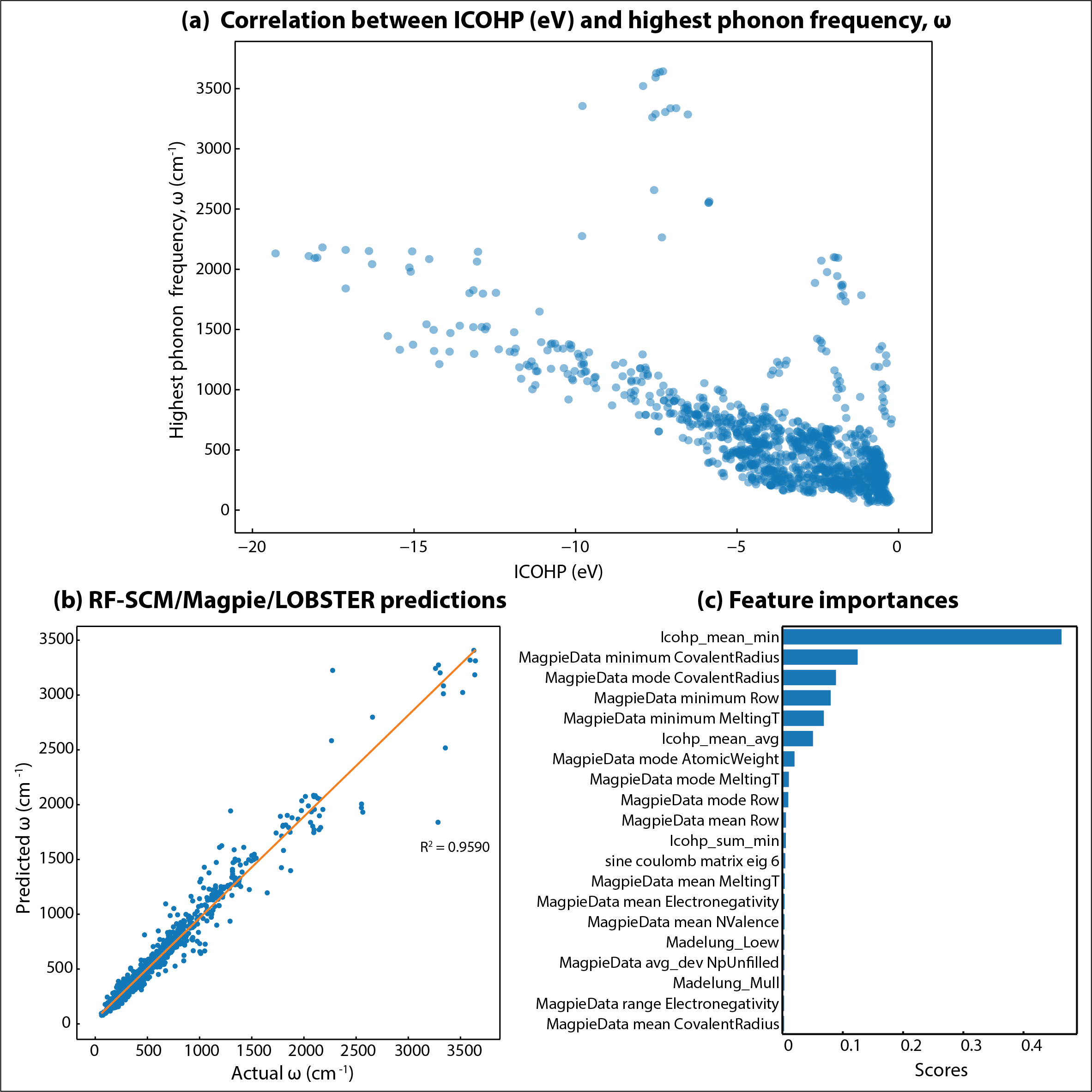}
  \caption{(a) The strongest ICOHP (eV) values plotted against the highest phonon frequency, $\omega(cm^{-1}$). (b) Predicted $\omega$ values from RF-SCM/Magpie/LOBSTER model for the whole dataset (c) Feature importance scores for RF-SCM/Magpie/LOBSTER model} 
 \label{fig:datautil}
\end{figure}

Lastly, we demonstrate the utility of our data by building a machine-learned model to predict the highest phonon frequency ($\omega$) as computed with harmonic phonon computations.\cite{petretto2018high} This property is also part of the Matbench benchmark set.\cite{dunn2020benchmarking} Therefore, a growing number of ML algorithms, such as MegNet\cite{chen2019graph}, ALIGNN\cite{choudhary2021atomistic}, MODNET\cite{de2021materials,de2021robust} have been used to predict the highest phonon frequency. We selected this property as ICOHP values (covalent bond strengths) have previously been correlated to force constants from harmonic phonon runs (e.g., in ref \cite{lderinger_vibrational_2015}) and should therefore be ideal features for harmonic phonon properties. Also, we have computed LOBSTER data for almost all the compounds included in the benchmark phonon dataset in the Matbench test suit\cite{dunn2020benchmarking}. We note that bonding analysis only requires a fraction of the computational time of typical phonon runs, as only one static DFT run and post-processing with Lobster are required. 
As a first step before developing the ML model, we checked linear correlations between our quantum-chemical bonding information and our target property. We found a clear correlation between the strongest ICOHP of each compound and the highest phonon frequency ($\omega$) (Fig.~\ref{fig:datautil}(a)). We can, however, see at least two different trends. We assume this is related to the fact that the highest phonon mode can stem from very different vibrations. Some might be pure stretching vibrations, and others could be collective vibrations involving all atoms. In the first case, mostly one specific bond and one specific ICOHP would have high importance for the phonon mode, whereas in the latter case, all interactions and, therefore, more than one ICOHP within the material would play a role. This observed correlation indicates that using LOBSTER data in ML studies as an additional feature could improve the predictive models.

To test this hypothesis, we first transform the data from summarized bonding information (including all types of bonds and not only cation-anion bonds) of the lightweight JSON files to features for our ML models. For this purpose, we developed a featurizer that accepts these JSON files as input and provides mean, min/max, standard deviation ICOHP values, and Madelung energies based on Mulliken and L{\"o}wdin as output in a tabular format for each compound. An explanation of the generated features is provided in Table.~\ref{tab:featurizer}. 
\begin{longtable}[c]{|l|l|}
\hline
\rowcolor[HTML]{2985F0} 
Features  & Description \\ \hline
\endfirsthead
\multicolumn{2}{c}%
{{\bfseries Table \thetable\ continued from previous page}} \\
\hline
\rowcolor[HTML]{2985F0} 
Feature name  & Description \\ \hline
\endhead
Icohp\_mean\_avg  & \begin{tabular}[c]{@{}l@{}}Average of all relevant ICOHPs per bond at symmetrically inequivalent sites in the structure\end{tabular} \\ \hline
Icohp\_mean\_max  & \begin{tabular}[c]{@{}l@{}}Maximum of all relevant ICOHPs per bond at symmetrically inequivalent sites in the structure\end{tabular} \\ \hline
Icohp\_mean\_min  & \begin{tabular}[c]{@{}l@{}}Minimum of all relevant ICOHPs per bond at symmetrically inequivalent sites in the structure\end{tabular} \\ \hline
Icohp\_mean\_std  &  \begin{tabular}[c]{@{}l@{}}Standard deviation of all relevant ICOHPs per bond at symmetrically inequivalent sites \\ in the structure\end{tabular} \\ \hline
Icohp\_sum\_avg  & \begin{tabular}[c]{@{}l@{}}Average of all relevant ICOHP sums at symmetrically inequivalent sites in the structure \end{tabular} \\ \hline
Icohp\_sum\_max  & \begin{tabular}[c]{@{}l@{}}Maximum of all relevant ICOHP sum at symmetrically inequivalent sites in the structure \end{tabular}\\ \hline
Icohp\_sum\_min  & \begin{tabular}[c]{@{}l@{}}Minimum of all relevant ICOHP sums at symmetrically inequivalent sites in the structure \end{tabular} \\ \hline
Icohp\_sum\_std  & \begin{tabular}[c]{@{}l@{}}Standard deviation of all relevant ICOHP sums at symmetrically inequivalent sites \\ in the structure \end{tabular} \\ \hline
bonding\_perc\_avg  & \begin{tabular}[c]{@{}l@{}}Average of bonding percentages below Fermi level from  COHPs at symmetrically \\ inequivalent sites in the structure \end{tabular} \\ \hline
bonding\_perc\_max  & \begin{tabular}[c]{@{}l@{}}Maximum bonding percentage below Fermi level from  COHPs at symmetrically \\ inequivalent sites in the structure \end{tabular} \\ \hline
bonding\_perc\_min  & \begin{tabular}[c]{@{}l@{}}Minimum bonding percentage below Fermi level from  COHPs at symmetrically \\ inequivalent sites in the structure \end{tabular} \\ \hline
bonding\_perc\_std  & \begin{tabular}[c]{@{}l@{}}Standard deviation of bonding percentages below Fermi level from  COHPs at symmetrically \\ inequivalent sites in the structure \end{tabular} \\ \hline
antibonding\_perc\_avg  & \begin{tabular}[c]{@{}l@{}}Average of anti-bonding percentages below Fermi level from  COHPs at symmetrically \\ inequivalent sites in the structure \end{tabular} \\ \hline
antibonding\_perc\_max  & \begin{tabular}[c]{@{}l@{}}Maximum anti-bonding percentage below Fermi level from  COHPs at symmetrically \\ inequivalent sites in the structure \end{tabular} \\ \hline
antibonding\_perc\_min  &\begin{tabular}[c]{@{}l@{}}Minimum anti-bonding percentage below Fermi level from  COHPs at symmetrically \\ inequivalent sites in the structure \end{tabular} \\ \hline
antibonding\_perc\_std  & \begin{tabular}[c]{@{}l@{}}Standard deviation of anti-bonding percentages below Fermi level from  COHPs at symmetrically \\ inequivalent sites in the structure \end{tabular} \\ \hline
Madelung\_Mull  & \begin{tabular}[c]{@{}l@{}}Madelung energy of the structure derived from Mulliken charges\end{tabular} \\ \hline
Madelung\_Loew  & \begin{tabular}[c]{@{}l@{}}Madelung energy of the structure derived from L{\"o}wdin charges \end{tabular} \\ \hline
\caption{ICOHP Features extracted using featurizer from LOBSTER Lightweight JSONS}
\label{tab:featurizer}\\
\end{longtable}

Such an approach is commonly used to generate material descriptors for machine learning of material properties.\cite{ward2018matminer,de2021materials} The authors would like to emphasize that the aim of this experiment is not to build the best predictive model but to demonstrate the influence of using LOBSTER data as features in ML studies. We assume that graph-based models which allow adding the bonding descriptors as edge features might be more predictive. That being said, to test the influence on a model's predictive performance, we trained and evaluated two Random forest (RF) regressor \cite{breiman2001random} models. Both models differ only in the input feature sets. RF-SCM/Magpie model consisted of SineCoulombMatrix\cite{faber2015crystal} and elemental Magpie\cite{ward2016general,ward2018matminer} features (mean, average deviation, range, and max/min statistics) as obtained from AutoFeaturizer module of Automatminer\cite{dunn2020benchmarking} with ``debug'' preset (180 features). The input feature set and a fixed set of 500 estimators for RF regressor match the matbench v0.1 RF-SCM/Magpie model.\cite{dunn2020benchmarking}. The input feature set of the RF-SCM/Magpie/LOBSTER model consisted of the identical feature space as the RF-SCM/Magpie model, and it was augmented by LOBSTER data obtained from our featurizer (199 features). We ensure the train and test sets used for evaluation are identical in both models by setting the same random state seed. The models are evaluated using the nested cross-validation (CV) approach. The inner five-fold CV is used only to optimize the feature selection algorithm (MultiSurfstar\cite{Urbanowicz2017Benchmarking}) hyperparameter, i.e., the number of features selected. The hyperparameters of the RF regressor are not tuned. The  CV statistics across all five test sets for both models are summarized in Table.~\ref{tab:mlmodelstat}. Our RF-SCM/Magpie model performs similarly to the one reported on the matbench test suit.\cite{dunn2020benchmarking} Including LOBSTER data as features in model input shows an apparent increase in model prediction accuracies. An increase in accuracies by approximately 27\% for mean absolute error (MAE), 28 \% for Max Errors, 32 \% for root mean squared errors (RMSE), and 5 \% for $R^2$ is observed. 
\begin{table}[h]
\centering
\resizebox{.97\textwidth}{!}{%
\begin{tabular}{|c|c|c|c|c|}
\hline
\rowcolor[rgb]{0.161,0.522,0.941} 
Model & MAE & Max Errors & RMSE & R$^2$ \\ \hline
RF-SCM/Magpie & 68.047 (± 7.502) & 1208.329 (± 380.017) & 149.611 (± 19.762) & 0.905 (± 0.027) \\ \hline
RF-SCM/Magpie/LOBSTER & 49.885 (± 1.941) & 866.373 (± 335.674) & 100.893 (± 9.160) &0.957 (± 0.012) \\ \hline
\end{tabular}%
}
\caption{Comparison of RF model accuracies across five-fold nested cross-validation test sets. The numbers in the parenthesis depict the standard deviation of the metrics. (MAE: Mean absolute error, RMSE: Root mean square errors, R$^2$: coefficient of determination)}
\label{tab:mlmodelstat}
\end{table}

On further analysis of the best-performing model (RF-SCM/Magpie/LOBSTER), it is found that the algorithm only needs 50 input features after feature selection for predicting the target values more accurately compared to RF-SCM/Magpie, where all 180 were required. This result demonstrates that significantly fewer features are needed when bonding-related features from LOBSTER are included as features. We looked at the feature importance scores readily available for RF models to further analyze the best model. As seen in Fig. ~\ref{fig:datautil} (c), the better performing  RF-SCM/Magpie/LOBSTER model shows that the \emph{'ICOHP\_mean\_min'} feature, which indicates the ICOHP value for the most covalent bond in a compound largely contributed to learning the target property of interest. This is the same feature that shows the high correlation in Fig. ~\ref{fig:datautil} (a). Shapley\cite{lundberg2020local} values computed for the RF models to assess the impact of input features on model prediction also show a similar trend (Plots are provided as part of the repository  \href{https://doi.org/10.5281/zenodo.7802318}{10.5281/zenodo.7802318}). This result further supports our hypothesis that including bonding-related features as material descriptors in ML studies of materials properties not only improves accuracies of predictions but also helps to understand the relationships between material properties and chemical bonding. Here, we clearly see a suspected relationship between covalent bond strengths and harmonic phonon properties.

\section*{Usage Notes}
In this work, we provided a Quantum-Chemical Bonding Database to predict and discover new materials. This database, at the moment, consists of summarized COHP-based bonding analysis information ready to be used for ML studies. It also includes (I)COOP, (I)COBI, DOS, atomic charges, and Madelung energies in the computational data JSON files. In addition, we also demonstrated a use-case scenario of how our data could be used for ML studies. This by no means implies that our data should be used in such a manner only. End users are encouraged to explore further.  


\section*{Code availability}
The following program versions have been used in this study: pymatgen 2023.1.9, atomate 1.0.3, LobsterPy 0.2.9, LOBSTER 4.1.0, and VASP 5.4.4. All the scripts used in this study, from starting the workflow, generating data records, reproducing technical validation plots, and ML model evaluations, can be accessed here: \href{https://github.com/naik-aakash/lobster-database-paper-analysis-scripts}{https://github.com/naik-aakash/lobster-database-paper-analysis-scripts} (\href{https://doi.org/10.5281/zenodo.7802318}{10.5281/zenodo.7802318}).

\section*{References}
\vspace{-0.5cm}

\section*{Data Citations}

\section*{Acknowledgements}
A.N. and J.G would like to acknowledge the Gauss Centre for Super computing e.V. (www.gauss-centre.eu) for funding this project by providing generous computing time on the GCS Supercomputer SuperMUC-NG at Leibniz Super computing Centre (www.lrz.de) (project pn73da). The authors thank Katharina Ueltzen for bringing to light an issue with supplied magnetic moments from INCAR not being read correctly during VASP static runs and for helping us rectify the same. A.N. thanks Franziska Emmerling and Manuel Kupper for their feedback on the manuscript in BAM’s MatChIngCamp. J.G. thanks Geoffroy Hautier and Matthew Horton for helpful discussions, and A.N. and J.G. thank Alex Ganose for reviewing the pydantic schema used in this study as part of our atomate2 pull request for a new Lobster workflow. We also acknowledge the maintainers of pymatgen.

\section*{Author contributions statement}
A.N. performed the high-throughput calculations and data collection with help from J.G. and P.B. C.E. performed additional computations to analyze the BaO$_2$ case. All authors analyzed the data. A.N., C.E., and J.G. wrote the manuscript with inputs from all authors. A.N. and J.G. have planned the study. A.N., N.D., P.B., and J.G. contributed to the ML model.  

\section*{Competing interests}
The authors declare no competing interests.
\newpage
\input{supplementary.tex}
\end{document}

%% file: supplementary.tex
\setcounter{figure}{0} 
\setcounter{equation}{0} 
\setcounter{table}{0} 
\setcounter{section}{0}
\renewcommand{\thefigure}{S\arabic{figure}}
\setcounter{figure}{0}
\makeatletter
\renewcommand{\thetable}{S\@arabic\c@table} 
\makeatother
\makeatletter
\renewcommand{\theequation}{S\@arabic\c@equation} 
\makeatother
\begin{center}
    \section*{\centering Supplementary information}
\end{center}
\noindent\textbf{Projected density of states (PDOS) benchmarking}
\begin{figure}[H]
  \begin{minipage}{\linewidth}
  \centering
  \includegraphics[width=.90\textwidth]{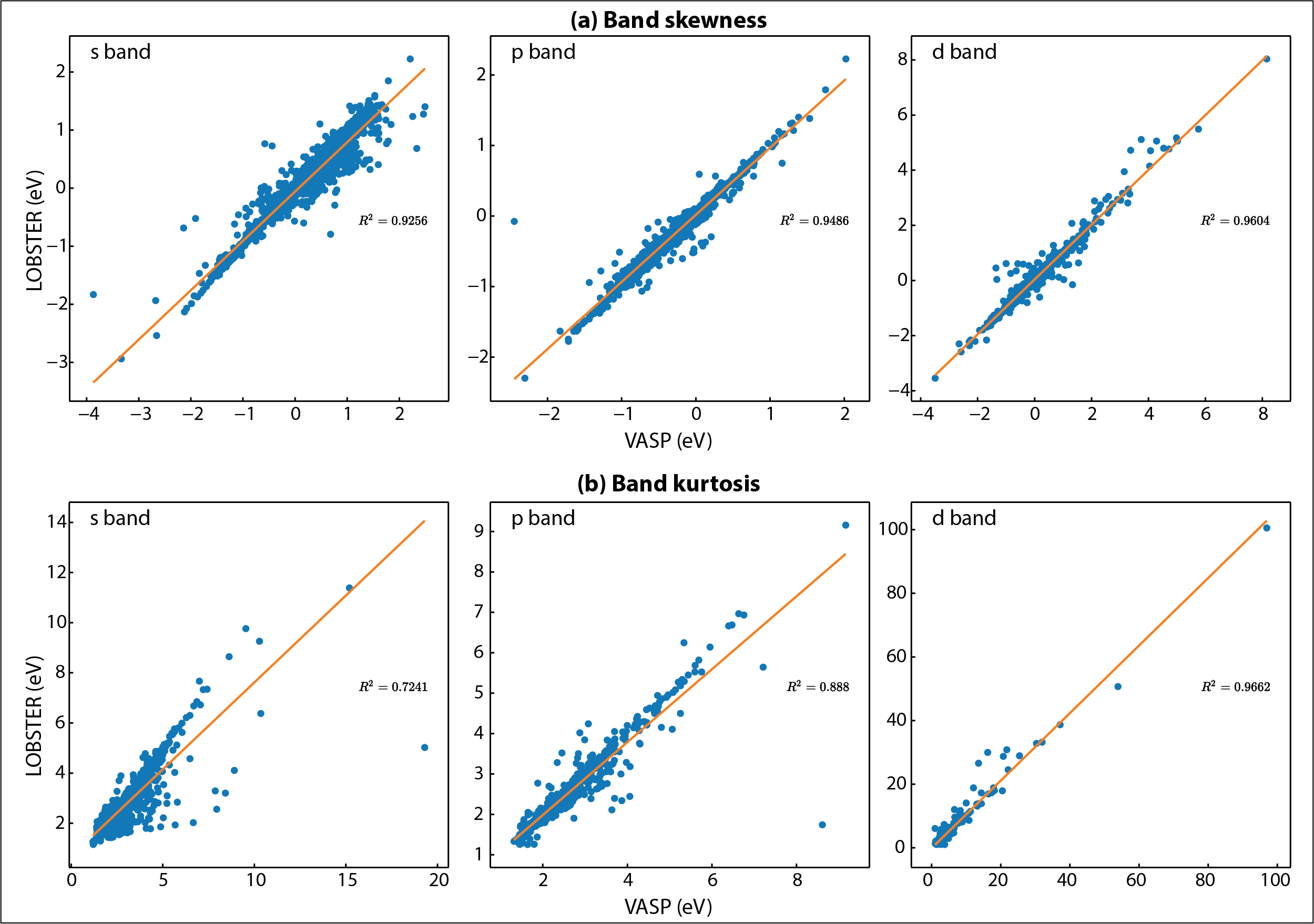}
  \end{minipage}
  \caption{(a) Band skewness and (b) band kurtosis comparison of projected DOS (s, p and d bands) for first non-zero energy range below Fermi level ($E_F$) obtained from LOBSTER and VASP runs. Both figures show that the projected DOS from LOBSTER runs are in reasonable agreement with our reference VASP data.}
 \label{fig:spdbandskewnkurt}
\end{figure}

\begin{figure}[H]
\centering
\includegraphics[width=.70\linewidth]{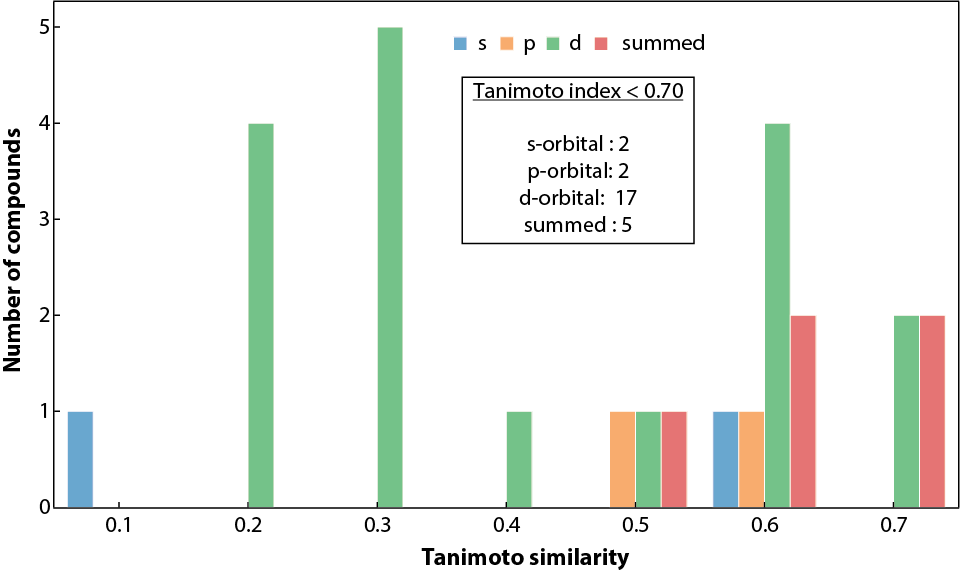}
\caption{Histogram of Tanimoto index (< 0.70) computed between VASP and LOBSTER.}
\label{fig:lowtanimoto}
\end{figure}

\begin{figure}[H]
\centering
\includegraphics[width=.95\linewidth]{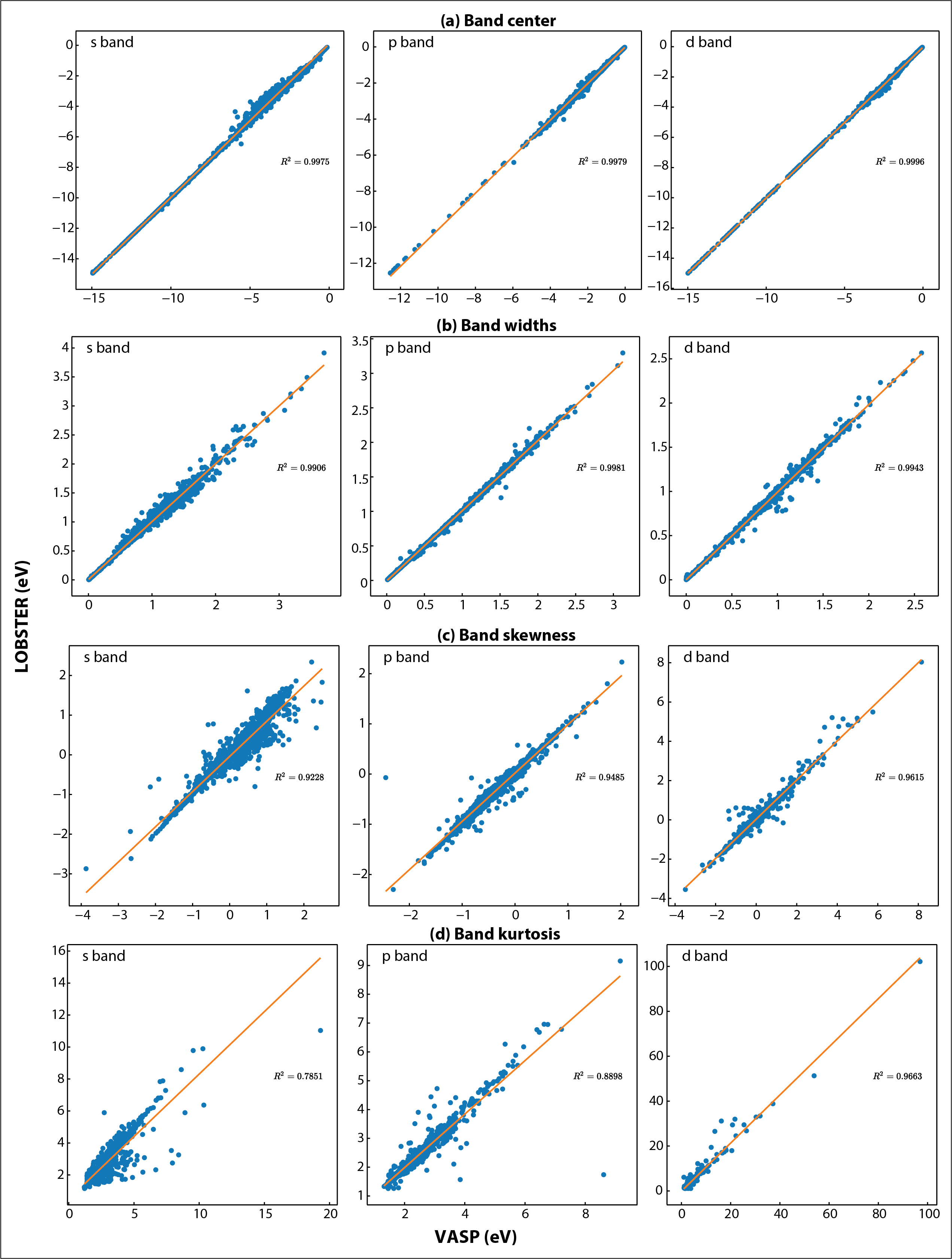}
\caption{(a) Band center and (b) width (c) skewness and (d) kurtosis comparison of projected DOS (s, p and d bands) for first non-zero energy range below the Fermi level ($E_F$) obtained from LOBSTER (non LSO) and VASP runs. }
\label{fig:nonlsobf}
\end{figure}

\noindent\textbf{Atomic charges and coordination environments}

\begin{figure}[H]
\centering
\includegraphics[width=.65\linewidth]{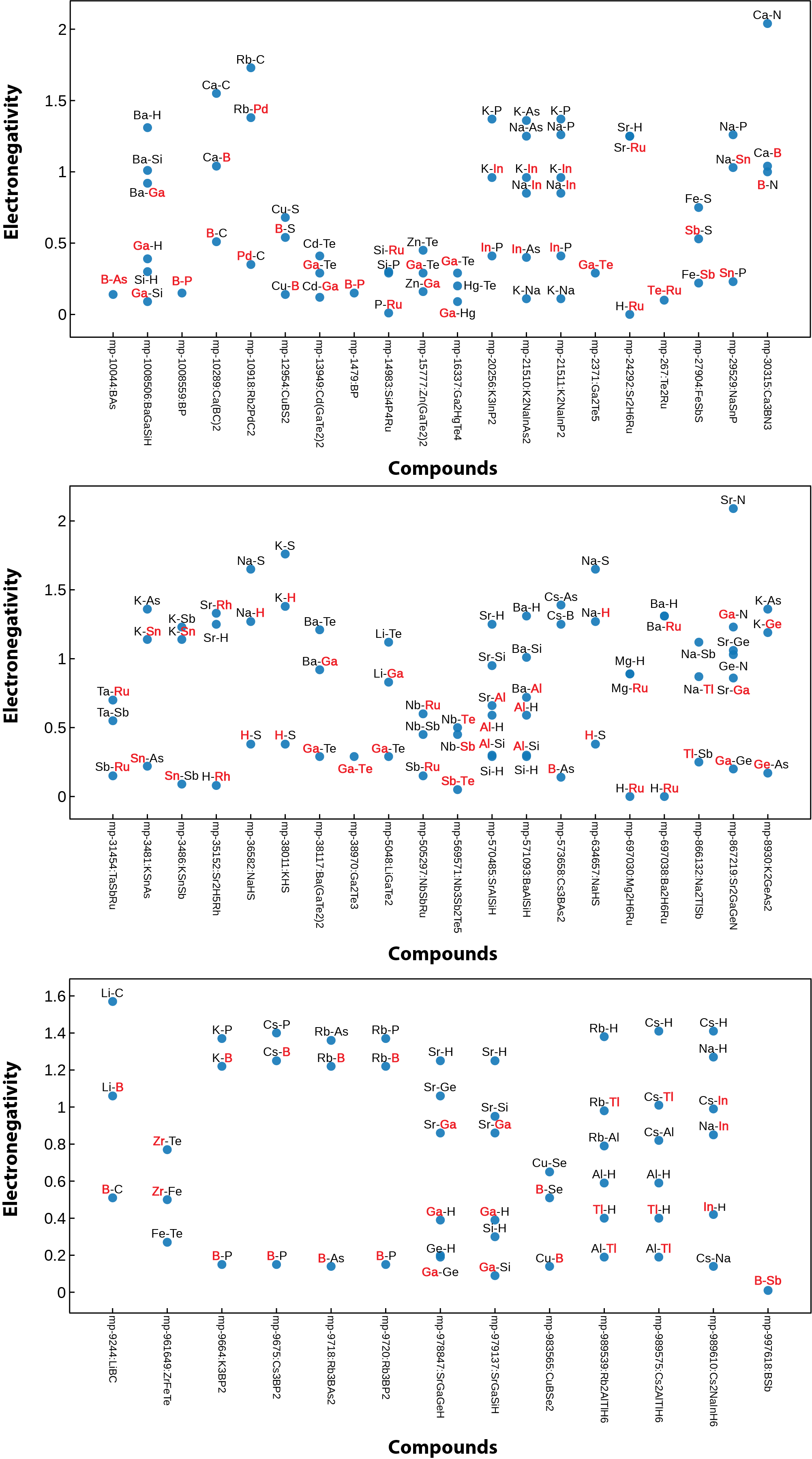}
\caption{Electronegativity differences scatter plot for the compounds where cations and anions assignment differs from LOBSTER and BVA methods. (Text annotations in RED depict the elements where cation-anion classification disagreements are observed).}
\label{fig:endiff}
\end{figure}

\begin{figure}[H]
\centering
\includegraphics[width=.80\linewidth]{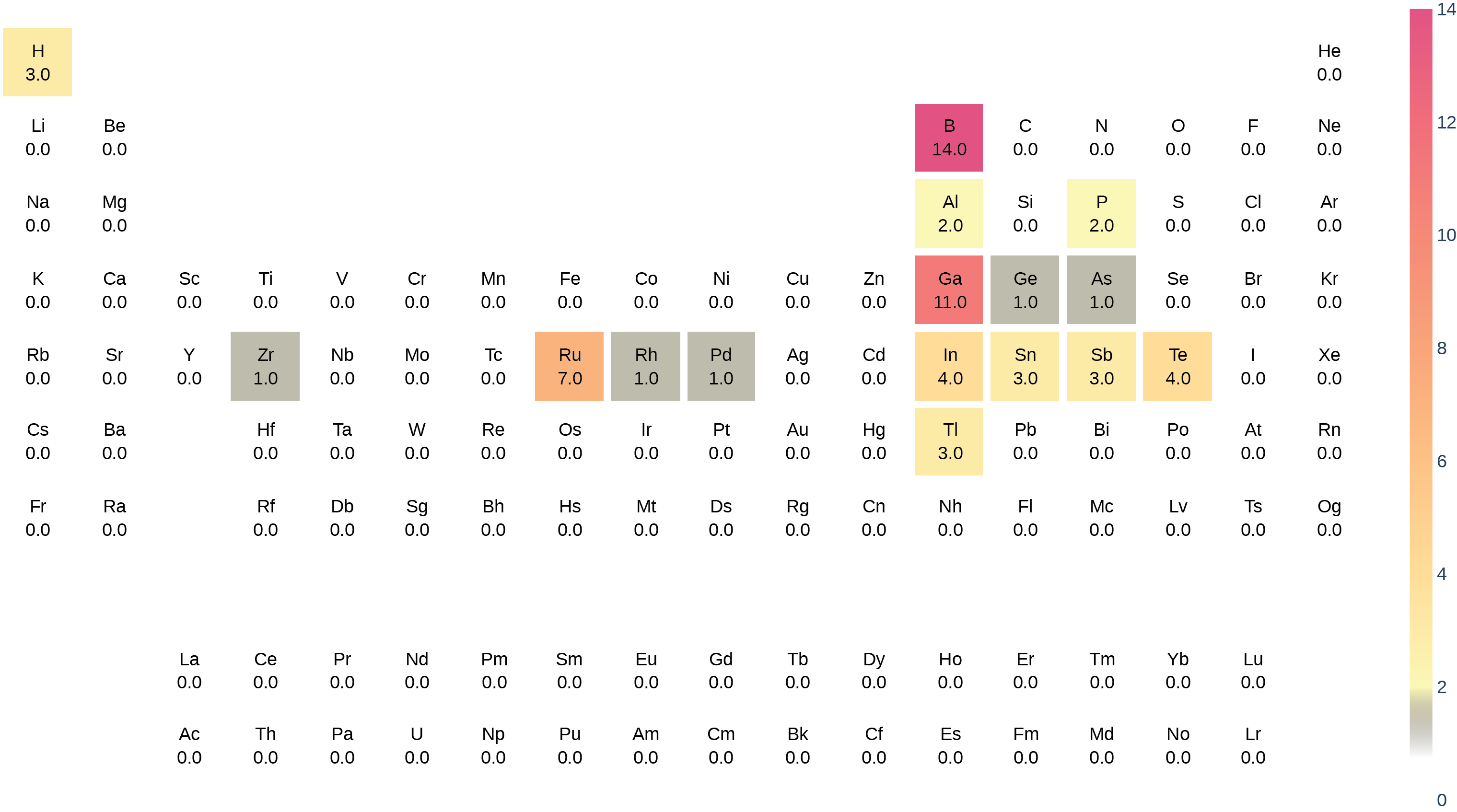}
\caption{Elements for which cations and anions assignment classification differs between LOBSTER and the BVA methods depicted in the form of a heatmap. The heatmap was plotted with pymatviz \cite{Riebesell_Pymatviz_2022}}
\label{fig:heatmap}
\end{figure}